# Characterization of Conventional Endovascular Devices in Treatment of Abdominal Aortic Aneurysms

**Yara Alawneh[1], James J. Zhou[1], Alykhan Sewani[1], Andrew Dueck[2], and M. Ali Tavallaei[1]**

[1]Faculty of Engineering and Architectural Science, Toronto Metropolitan University, Toronto, Ontario, Canada
[2]Department of Vascular Surgery, Sunnybrook Health Sciences Centre, Toronto, Ontario, Canada

Corresponding author: M. Ali Tavallaei (e-mail: ali.tavallaei@torontomu.ca).

This work was supported in part by the Canada Research Chairs Program (CRC-2019-00012), the Natural Sciences and Engineering Research Council of Canada (NSERC), the Canadian Foundation for Innovation (CFI), the Ontario Government, as well as Toronto Metropolitan University.

**ABSTRACT** Abdominal Aortic Aneurysms (AAA) are often repaired through an Endovascular approach known as EVAR. The success and duration of these challenging procedures are primarily attributable to the accuracy and reliability of navigating corresponding interventional devices. This study investigates the performance of conventional non-steerable and steerable catheters in endovascular aneurysm repair (EVAR) procedures, focusing on two primary metrics: reachable workspace and gate cannulation success. We developed two abdominal aortic aneurysm (AAA) phantoms using patient CT images for our experiments. Under X-ray fluoroscopy guidance, the reachable workspace was quantified, and gate cannulation success rates, cannulation time, and fluoroscopy times were recorded for both non-steerable and steerable catheters and were compared. We were unable to observe statistically significant differences between the two catheter types in overall cannulation success rates or fluoroscopy time. However, in challenging anatomical scenarios (particularly a more challenging gate location), the steerable catheter showed statistically significant advantages in success rates and cannulation times. While there were no statistical differences in reachable workspace between non-steerable and steerable catheters when considering the whole aneurysm, segmented analysis showed that the steerable catheter performed better in the central region, and non-steerable catheters performed better in the peripheral region. This study provides a systematic method for quantifying the performance of endovascular devices. The findings suggest that while steerable catheters may offer advantages in complex anatomical conditions, non-steerable catheters are preferable in peripheral areas of the aneurysm. These insights can inform catheter selection in EVAR, potentially influencing device design and clinical practice.

## I. INTRODUCTION

Abdominal Aortic Aneurysms (AAAs) affect roughly 4.8% of adults worldwide; this risk extends to approximately 20% for males older than 75 [1], [2]. They result from the overexpansion of the abdominal aorta, typically spanning from the common iliac arteries to just below the renal arteries, to greater than 3.0 cm in diameter due to the weakening of the artery wall [3], [4]. As the aneurysm grows and ages, the risk of rupture increases [3]. Upon rupture, there is a 65-85% mortality rate [4]. Therefore, it is imperative to depressurize the weakened vessel wall or aneurysmal sac.

Treating AAAs requires the implantation of an endograft, which forms the new aortic wall. If the procedure is successful, it leads to a reduction of pressure on the aneurysm and mitigates the risks of rupture. Implanting an endograft can be achieved through open surgery or an endovascular aneurysm repair (EVAR), with vascular surgeons preferring EVAR due to a lower perioperative mortality rate [3], [5].

Under x-ray guidance, EVAR involves using a catheter-based delivery system to deploy a stent-graft that acts as the new vessel wall for the abdominal aorta. These stent-grafts typically have a large lumen (for the aorta) that splits into two smaller lumens designed to bifurcate into the left and right common iliac arteries. With the stent-graft partially deployed by the delivery system from the ipsilateral limb, the surgeon uses a guidewire and catheter combination to perform a gate cannulation from the contralateral limb, where they guide the wire into the opposite lumen of the stent-graft [5]. Upon successful gate cannulation, the surgeon can proceed to implant a covered stent to bridge the stent-graft from the aorta into the common iliac. This procedure, when successful, isolates and depressurizes the aneurysm sac [4], [5].

Although EVAR has a relatively high success rate (95%) when compared to other catheter-based procedures [6], [7] these procedures take 2-4 hours on average to perform [8]. Prolonged procedures lead to increased exposure to ionizing radiation for the patients but also risk various other complications such as respiratory depression, contrast-induced nephropathy, cognitive issues, tissue or vessel damage, etc.[9], [10]. For the clinicians and staff, the prolonged procedures are also a hazard. They are exposed to ionizing backscattered radiation while also suffering from chronic back and neck pain due to the need to wear heavy lead aprons for radiation protection for prolonged periods.

Apart from the long procedure duration, there is also further inefficiency in the procedure outcomes. Up to 40% of EVARs fail over time [11]–[13], with reported complication rates ranging from 16% to 30%[14]. These complications and failures are usually related to the stent-graft (e.g., endo leaks); however, there is still a large subset of complications related to the operation of the catheters[12]. Given the post-operative failure rate, these complications and failures can compound, leading to higher reintervention rates and higher costs for the patient and healthcare system. With these inefficiencies in procedure duration and outcome, we must understand the root causes for such limitations and characterize them. Quantifiable measures of device performance will also permit the assessment of new solutions aimed at addressing the limitations of existing tools.

We believe one of the main causes of inefficiencies in conventional procedures can be traced to the steering and navigation of non-steerable catheters and tendon-actuated steerable catheters used to manipulate the guidewire as part of these procedures. These long and flexible devices are manually operated from outside of the patient body. EVAR requires accurate and reliable navigation and control of the device tip relative to the anatomy to navigate and position the device's tip accurately, for example, for contralateral gate cannulation. Depending on the patient's anatomy, this can be extremely challenging [15]–[17]. Two contributors to this limitation include: 1) the soft and compliant nature of the devices create unpredictable interactions with the anatomy along their path, which restricts the accessible workspace and control of the catheter tip; and, 2) the lack of 3D anatomical information and lack of depth perception in conventional 2D projection x-ray imaging used to guide these procedures. Ultimately, in the context of an EVAR, these limitations can lead to inefficiencies in reliable and accurate device steering and navigation, which can be associated with prolonged procedures and suboptimal graft positioning and corresponding complications.

To further understand this problem and target the root cause for challenging catheter control, researchers have developed models to explain how the remote catheter inputs transfer into catheter tip motion, the key to successful catheter-based procedures. For example, in previous work [18] [19], the performance of conventional devices for applications of revascularization has been characterized using phantom models. The results demonstrated that conventional devices have major limitations in reachable workspace and force delivery, ultimately limiting the potential utility for their specific intended task or revascularization. In other work, Yu et al. developed a model for how steerable catheters interact with the anatomy[20]. Another group, Jones et al., simulated the accessible workspace of four commonly used endovascular catheters and used their models to evaluate off-label catheter modifications[21]. While other groups have aimed at developing phantoms mimicking AAA to simulate or replicate a procedure [22]–[28], we are not aware of any work that aims to quantifiably characterize device performance within such phantom models, with performance metrics related to the success of the procedure and reachable workspace and to use them to compare existing conventional devices objectively.

Therefore, this study aims to quantify the performance of conventional non-steerable and steerable catheters used during infrarenal EVAR, with two prime metrics of reachable workspace and gate cannulation, using AAA phantom models.

## II. METHODS

We evaluated two classes of conventional catheters, non-steerable catheters (**N-SCs**) and steerable catheters (**SCs**), in a set of 3D-printed AAA phantoms under x-ray fluoroscopy guidance..

During the gate cannulation, the operator must navigate a guidewire with the help of a catheter from the common iliac artery, into the aneurysmal sac, and the contralateral gate. This becomes more challenging with complex anatomy (e.g., high tortuosity in the iliac) [29]. The procedure requires accurate guidewire steering via a catheter and the ability to steer within a large lumen, all while guided with x-ray fluoroscopy. The AAA CT scans used to create the 3D-printed phantom models were selected to replicate these key challenges based on clinician input.

*Experiment Overview*

In these phantom experiments, we asked two novice users with one year of experience in catheter testing, one intermediate user, a catheter engineer with over 15 years of experience evaluating catheters, and one experienced interventionalist with over 15 years of experience to participate. We compared N-SCs with SCs in performing two tasks: **1)** a contralateral gate cannulation of a stent-graft (Endurant, Medtronic, Dublin, Ireland) and **2)** a reachable workspace analysis. We designed these experiments to quantify the success rates and procedure times required to access the contralateral gate and the ability of the catheter to navigate effectively within the large aneurysm. Paired with each catheter, users could interchangeably use two guidewires, a straight or a J-shaped guidewire (Glidewire®, Terumo Interventional Systems, Japan).

*Phantom Overview*

To create the phantoms, we obtained a set of stereolithography files from CT scans that met the criteria for EVAR (diameter greater than 5.0 cm) and based on input from our expert interventionalist. We modified the files using Meshmixer to allow 3D printing and use for each experiment (Formlabs Clear Resin, Formlabs, Somerville, MA, USA). For the first experiment of contralateral gate cannulation, the 3D-printed phantom was segmented along the coronal plane to allow easy placement and removal of the endograft (Fig. 1). This allowed us to deploy the endograft in different challenging orientations and then close the aneurysm by fastening the removable section of the phantom.

For experiment 2, the workspace analysis, we 3D printed a phantom segmented axially across the midline of the aneurysm. Next, we designed and 3D printed a backing plate to hold modeling clay and position at the aneurysm's midline. As the contact of the wire and the clay leaves a visible trace, it allows for characterizing and quantification of the device's reached workspace.

*A. Experiment 1: Contralateral Gate Cannulation*

Using the phantom designed for experiment 1, we placed the endograft inside the aneurysm in two different positions, one representing an ideal endograft deployment in the contralateral quadrant of the aneurysm (Fig 3. A) and the other representing a challenging deployment, in our case, in the dorsal quadrant of the aneurysm (Fig 3. B, C). Blinded to the orientation of the stent graft, users had 10 minutes to access the contralateral gate under fluoroscopy using a non-steerable catheter and a steerable catheter.

Based on their discretion, users could switch between the straight and J-shaped guidewire for the N-SC attempts. They also had four different N-SCs. These catheters were the 4Fr KMP GLIDECATH® (Terumois, Japan), the 5Fr Impress Angiographic Catheter CB2 (Merit Medical, UT, USA), the 5Fr Impress Angiographic Catheter Vector 5 (Merit Medical, UT, USA), and the 5Fr Soft-Vu SoS Omni Selective 1 (AngioDynamics, NY, USA).

Operators used the devices in the following order: **KMP, CB2, Vector 5, and SoS Omni** as recommended by our expert interventionalist. If the users failed to cannulate the gate with a catheter (e.g., the KMP catheter), we recorded it as a failure, recorded the radiation dosage, and recorded a procedure time of 10 minutes. Only after failure do users move on to the subsequent non-steerable catheter. If they succeeded, we recorded it as a success and the radiation dosage and actual procedure time were also recorded; they would not use the subsequent catheter in the sequence. Each of these attempts is independent of each other with their own success or failures, radiation dosages, and total procedure times. We grouped the results from all the N-SCs together. Ultimately, this meant at least 4 data points for each gate cannulation position.

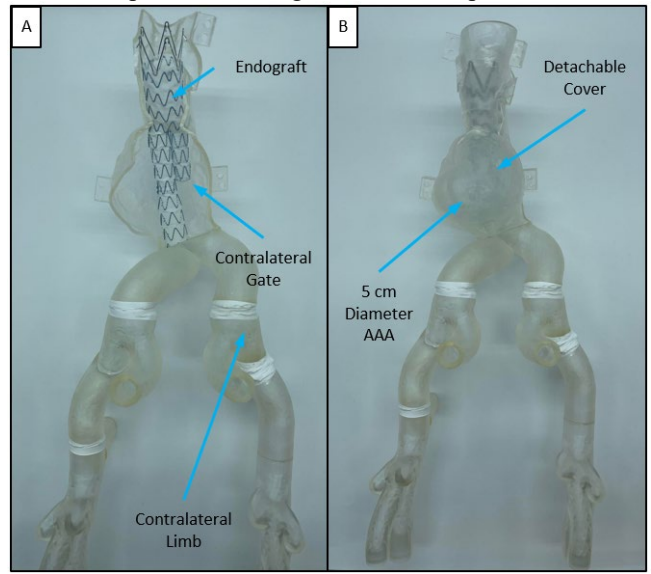

**Figure 1.** .A. segmented aneurysm showing endograft placement and contralateral gate for cannulation. B. Complete phantom assembly.

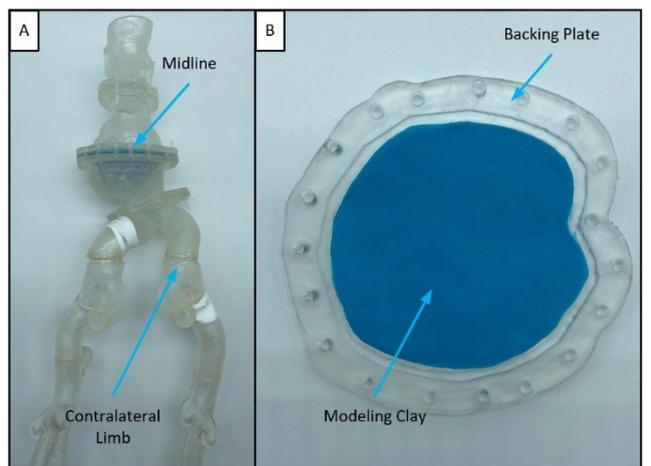

**Figure 2.** *A. Phantom for workspace analysis. B. Backing plate with modeling clay to track the workspace of each catheter.*

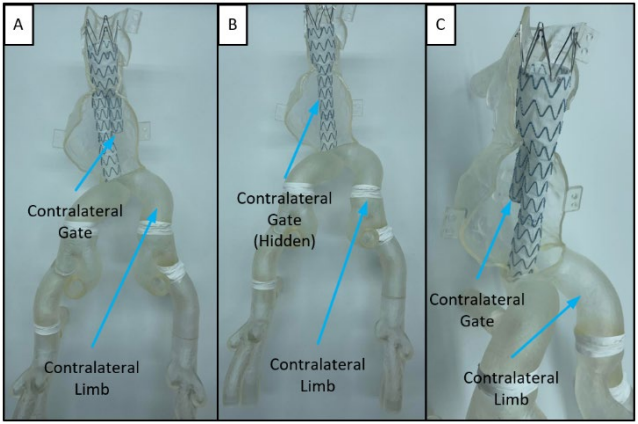

**Figure 3.** *A. Coronal view of position 1. B. Coronal view of position 2. C. Oblique view of position 2.*

For the SC attempts, with the two different guidewires, operators only used an 8.5Fr steerable catheter (Agilis™ NxT 8.5 Fr, Abbott Laboratories, Chicago, IL, USA). We recorded success rates of gate cannulation, cannulation times, and radiation dosages. Failure in gate access within the allotted time corresponds to recording maximum radiation dosage (proportional to 10 minutes of fluoroscopy time) and cannulation time (i.e., 10 minutes).

For both positions (shown in Fig. 3), using Fisher's exact tests ($\alpha = 0.05$) we compared the success rates of cannulation using both catheters (i.e., N-SC and SC). Next, we compared the cannulation times for all the positions using a two-tailed t-test ($\alpha = 0.05$). We then used a two-way ANOVA to evaluate the effect of the catheter and the gate position ($\alpha = 0.05$); we followed with post-hoc tests using Sidak's multiple comparisons to compare the cannulation times of each catheter for each position. Finally, we compared the radiation dosages for all positions using a two-tailed t-test ($\alpha = 0.05$). We then used a two-way ANOVA ($\alpha = 0.05$) to evaluate the effect of the catheter and the gate position on the radiation usage; we followed with Sidak's multiple comparisons to compare the dosages for each catheter in each position.

### B. Experiment 2: Workspace Analysis

Using the phantom designed for experiment 2, we moulded clay into a backing plate to track the workspace of the N-SCs while navigating from the contralateral limb (Fig. 2). For this experiment, the operators only used the straight guidewire. We also asked the same four operators to participate.

For the N-SCs, the users had 2 minutes per catheter (KMP, CB2, Vector 5, SoS Omni), or 6 minutes in total, to trace as much of the area as possible. We superimposed each catheter's workspaces to show each catheter's relative benefit and to determine the total impact of using all the N-SCs. For the SC, the users had 4 minutes to trace as much of the area as possible. The allocated time was based on the input from our experienced interventionalist, as it was assumed that 4 minutes is ample time for the SC to cover its reachable workspace.

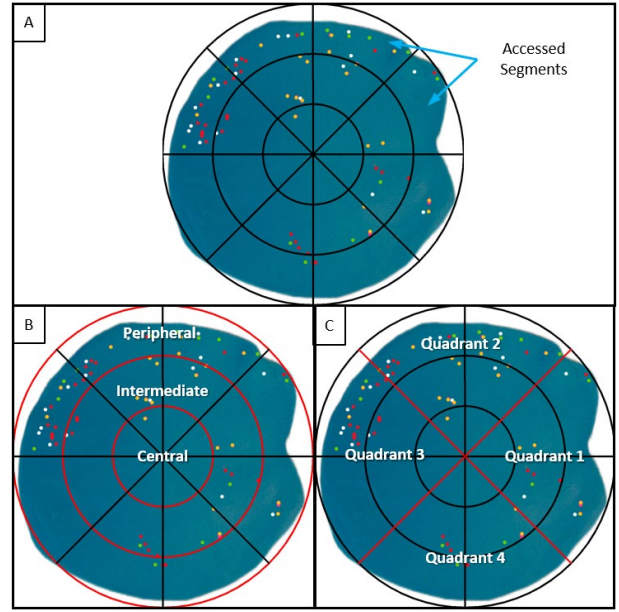

**Figure 4.** *A. Total workspace breakdown. B. Workspace segments grouped by radius. C. Workspace segments grouped by quadrant. The colors represent one user's attempts with the different catheters (i.e., KMP, CB2, Vector 5, SoS Omni).*

To enable a detailed assessment of the reachable workspace, we divided the workspace into 24 segments – 3 different radii and eight different angular sections – and calculated the percentage of segments accessed. We based the different radii segmentation such that the difference between each subsequent section is the approximate size of the median contralateral gate of the stent-graft (~12 mm), and we selected the radii as an important segment variable due to our previous phantom study comparing a N-SC with a SC in a peripheral artery phantom indicating that the catheters may preferentially navigate to a certain radius [19]. We divided the full circle into eight different angular segments. This number was chosen to allow for differentiation between various angles while avoiding redundancy. This segmentation allowed us to perform the following three analyses of the reachable workspace :

1. Total percentage of 24 segments covered (Fig. 4A)
2. Percentage of segments covered in the central region, the intermediate region, and the peripheral region (8 segments per region grouped by the three different radii, Fig 4. B)
3. Percentage of segments covered in each of the four quadrants: quadrant 1 (contralateral), quadrant 2 (ventral), quadrant 3 (ipsilateral), and quadrant 4 (dorsal). This amounted to 6 segments per quadrant (Fig 4. C).

We started by comparing the performance of all the N-SCs with the SC. Firstly, we compared the total percentage of segments covered between the N-SCs and the SC using a two-tailed t-test ($\alpha = 0.05$). Secondly, we compared the percentage of segments covered for the regions as grouped by radius. We used a two-way ANOVA with the first variable as the catheters and the second variable as the region ($\alpha = 0.05$);

TABLE I
GATE CANNULATION RESULTS SUMMARY

|  | N-SCs | SC | p-value |
|---|---|---|---|
| **Success Rate** | | | |
| Position 1 | 71.4% (5/7) | 50.0% (2/4) | 0.58 |
| Position 2 | 41.7% (5/12) | 100% (5/5) | 0.017* |
| Total | 52.6% (10/19) | 77.8% (7/9) | 0.25 |
| **Cannulation Time (mean ± S.D.) [s]** | | | |
| Position 1 | 297 ± 263 (n=7) | 372 ± 270 (n=4) | 0.83 |
| Position 2 | 460 ± 207 (n=12) | 176 ± 129 (n=5) | 0.047* |
| Total | 400 ± 236 (n=19) | 263 ± 215 (n=9) | 0.15 |
| **Radiation Dose (mean ± S.D.) [mGy]** | | | |
| Position 1 | 0.47 ± 0.51 (n=7) | 0.74 ± 0.51 (n=4) | 0.52 |
| Position 2 | 0.77 ± 0.50 (n=12) | 0.36 ± 0.28 (n=5) | 0.22 |
| Total | 0.65 ± 0.52 (n=19) | 0.53 ± 0.42 (n=9) | 0.54 |

we followed this with post-hoc tests using Sidak's multiple comparisons to understand the differences between the N-SCs and the SC for each region ($\alpha = 0.05$). Finally, we compared the percentage of segments covered for the regions as grouped by quadrants with the same statistical analysis.

We finished our analysis by comparing the performance of each N-SC (KMP, CB2, Vector 5, SoS Omni) and the SC. Firstly, we compared the total percentage of segments covered using a one-way ANOVA ($\alpha = 0.05$); we followed this with post-hoc tests using Tukey's multiple comparisons ($\alpha = 0.05$). Secondly, we compared the percentage of segments covered for the radial regions using a two-way ANOVA test ($\alpha = 0.05$), followed by Sidak's multiple comparisons. Finally, we compared the percentage of segments covered within each quadrant using a two-way ANOVA ($\alpha = 0.05$), followed by Sidak's multiple comparisons.

## III. RESULTS

### A. Experiment 1: Contralateral Gate Cannulation

*Table 1* shows a detailed summary of our measurements. In terms of our success rates, we found no statistical differences between the N-SC and the SC when comparing all the attempts (p=0.25, Fisher's exact test). However, the greater success rate when using the SC at 77.8% compared to the N-SC at 52.6% suggests that there may still be a benefit of using a SC. We again found no statistical differences when we analyzed the success rates of attempts when the gate resided in position 1 (within the contralateral quadrant). On the other hand, in position 2 (within the dorsal quadrant, our more challenging position), the SC statistically outperformed the N-SC (p=0.017, Fisher's exact test).

We found a similar trend in our cannulation times. We could not observe statistically significant differences when comparing all the attempts (p=0.15, two-tailed t-test). After including our position variable in our analysis, we found a statistically significant difference in cannulation time for position 2 in favour of the SC (p=0.047, Sidak's multiple comparisons). Despite radiation dosage being approximately proportional to the cannulation times, we found no statistical significance between any of our measurements with regard to dosage. However, there was still a trend towards reducing radiation usage with the SC in position 2.

Overall, this may suggest that when the contralateral gate resides in a more challenging position, the added control provided by the SC may offer an improvement, especially in success rate and cannulation time. However, this added control may not always be necessary based on the position of the gate. Perhaps in simpler anatomies and simpler endograft deployments, the reduced cost of an N-SC may outweigh the benefits of a SC.

### B. Experiment 2: Workspace Analysis

In our workspace analysis, when comparing the total percentage of segments covered between all N-SCs and the SC, we were unable to observe a statistically significant difference (p=0.44, two-tailed t-test). We found a statistically significant difference after dividing the aneurysm cross-section into three sections by radius (central, intermediate, and peripheral) (p=0.0026, two-way ANOVA). Our post-hoc tests comparing the N-SCs and the SC in the three different sections showed a significant difference in favour of the SC in the central region (p=0.0086), no observable significant difference in the intermediate region (p=0.22), and a significant difference in favour of the N-SCs in the peripheral region (p=0.0048). This suggests that the SC could potentially gain better access to a more centrally located contralateral while the N-SCs may better access a peripherally located contralateral gate.

However, some regions are less likely for the contralateral gate to reside upon initial deployment. So, we performed the same analysis by grouping the sections into quadrants. To reiterate, quadrant 1 is the quadrant on the contralateral side, quadrant 3 is on the ipsilateral side, and quadrants 2 and 4 are ventral and dorsal, respectively. We did not observe statistically significant differences between the N-SCs and the SC (p=0.42, two-way ANOVA), but we did observe a difference within the quadrants (p=0.002, two-way ANOVA). As we are concerned about the differences between the catheters within each quadrant, we further performed post-hoc tests that were unable to identify statistically significant differences (p>0.25). See Table 2 for a detailed summary of our results.

As we had grouped all five of our N-SCs together for the initial workspace analysis, we then went on to compare each specific N-SC catheter head-to-head to the SC. We observed statistically significant differences amongst the catheters for the total segments (p=0.0003, one-way ANOVA). Further post-hoc tests revealed differences between the KMP and the SC (p=0.0001), the Vector5 and the SC (p=0.0029), the CB2 and the SC (p=0.0204), and the SoS and the SC (p=0.0147), all in favour of the SC. This highlights a potential benefit of using a SC compared to the N-SCs.

TABLE II
N-SCs AND SC AREA PERCENTAGE COVERAGE BY SEGMENT

| Segments | N-SCs | SC |
|---|---|---|
| Centre | 25.0 ± 10.2 % | 65.7 ± 18.8 % |
| Intermediate | 59.4 ± 21.3 % | 81.3 ± 16.1 % |
| Peripheral | 59.4 ± 18.8 % | 15.6 ±12.0 % |
| Quadrant 1 | 25.0 ± 31.9 % | 54.2 ± 25.0 % |
| Quadrant 2 | 91.7 ± 9.6 % | 66.7 ± 19.2 % |
| Quadrant 3 | 50.0 ± 23.6 % | 50.0 ± 13.6 % |
| Quadrant 4 | 25.0 ± 28.9 % | 45.8 ± 8.3 % |
| Total | 47.9 ± 11.0 % | 54.2 ± 10.2 % |

TABLE II
PERCENTAGE AREA COVERAGE FOR EACH SPECIFIC CATHETER PER SEGMENT

| Segments | KMP | CB2 | Vector 5 | SoS | SC |
|---|---|---|---|---|---|
| Centre | 0.0 ± 0.0 | 18.8 ± 12.5 | 6.3 ± 7.2 | 9.4 ± 12.0 | 65.7 ± 18.8 |
| Intermediate | 9.4 ± 18.8 | 43.8 ± 7.2 | 25.0 ± 17.7 | 34.4 ± 6.3 | 81.3 ± 16.1 |
| Peripheral | 37.5 ± 20.4 | 34.4 ± 21.3 | 46.9 ± 6.3 | 50.0 ± 10.2 | 15.6 ±12.0 |
| Quadrant 1 | 8.3 ± 16.7 | 16.7 ± 23.6 | 8.3 ± 16.7 | 12.5 ± 16.0 | 54.2 ± 25.0 |
| Quadrant 2 | 16.7 ± 13.6 | 75.0 ± 16.7 | 50.0 ± 27.2 | 58.3 ± 21.5 | 66.7 ± 19.2 |
| Quadrant 3 | 29.2 ± 8.3 | 33.3 ± 13.6 | 41.7 ± 16.7 | 33.3 ± 13.6 | 50.0 ± 13.6 |
| Quadrant 4 | 8.3 ± 16.7 | 4.2 ± 8.3 | 4.2 ± 8.3 | 20.8 ± 21.0 | 45.8 ± 8.3 |
| Total | 15.6 ± 12.4 | 32.3 ± 5.2 | 26.0 ± 8.6 | 31.3 ± 4.2 | 54.2 ± 10.2 |

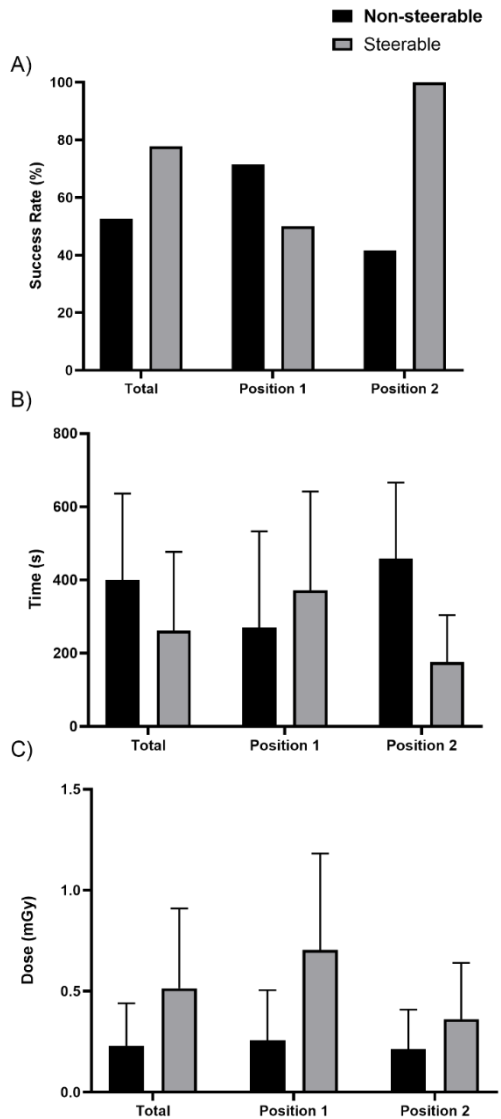

Figure 5. A. Success rates for gate cannulation. B. Total procedure times. C. Total radiation usage

While we were unable to observe statistically significant differences between the combined N-SCs grouped together with the SC, we were able to observe that a single SC significantly outperforms each single N-SC when compared directly. This implies that despite the added cost of a SC, there may be advantages to using a SC instead of multiple N-SCs for this application.

To further understand the performance of each catheter at different radial positions, we compared the catheters in the three radial groupings (central, intermediate, and peripheral).

We found statistical differences between the catheters and the different radii (p<0.0001, two-way ANOVA). We found no statistical differences amongst the N-SCs in our post-hoc tests in any of the regions. However, when compared to the SC in the central region, we found statistical differences between the KMP and the SC (p<0.0001), the CB2 and the SC (p=0.0002), the Vector5 and the SC (p<0.0001), and the SoS and the SC (p<0.0001), all in favour of the SC. In the intermediate region, we found statistical differences between the KMP and the SC (p=0.0087), the CB2 and the SC (p<0.0001), the Vector5 and the SC (p<0.0001), and the SoS and the SC (p=0.0002), all in favour of the SC. Finally, in the peripheral region, we only found differences between the Vector5 and the SC (p=0.0207) and between the SoS and the SC (p=0.0087), this time, in favour of the N-SCs. This echoes the comparison between all of the N-SCs and the SC. While the SC seems to more easily access centrally located contralateral gates, the N-SC group seems to more easily access peripherally located contralateral gates.

Finally, we analyzed the quadrant effect on catheter performance. Again, we found statistical differences between the catheters and the quadrants (p<0.0001, two-way ANOVA). In quadrant 1, our post-hoc tests revealed differences between the KMP and the SC (p=0.0032), the Vector 5 and the SC (p=0.0032), the CB2 and the SC (p=0.0237), and the SoS and the SC (p=0.009), all in favour of the SC. In quadrant 2, our post-hoc tests showed the KMP outperforming the CB2 (p<0.0001), the KMP outperforming the SoS (p=0.009), and the KMP outperforming the SC (p=0.0010). In quadrant 3, our post-hoc tests were unable to observe statistically significant differences. Finally, in quadrant 4, we observed significant differences between the KMP and the SC (p=0.0237), the Vector5 and the SC (p=0.009), and the CB2 and the SC (p=0.009), in favour of the SC. This may partially explain the differences in the cannulation times since the two conditions tested had the contralateral gate in quadrants 1 and 4. See Table 3 for a detailed summary of our results, and see Fig 6. for an example of our comparison.

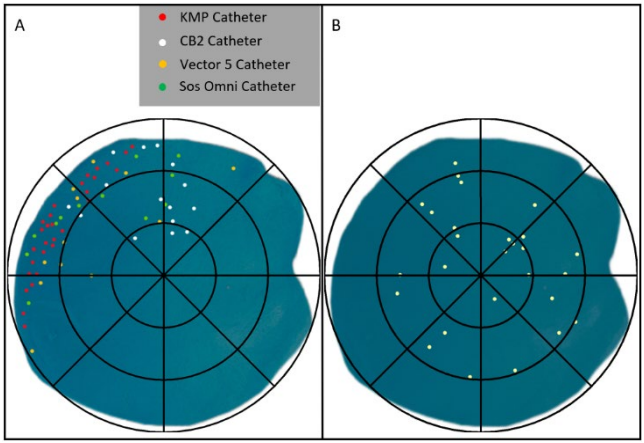

Figure 6. *Example workspace of non-steerable catheters (A) and steerable catheter (B) from one user..*

## IV. DISCUSSION

This study quantifies the performance of conventional non-steerable and steerable catheters used in minimally invasive procedures in the context of EVAR. One of the crucial steps during these procedures is the contralateral gate cannulation; the operator must navigate a guidewire into the short leg of the endograft using a catheter. Depending on the anatomy, accessing the contralateral gate can be challenging, requiring exceptional accuracy [17], [26]. Our study demonstrates these challenges in terms of cannulation success, cannulation times, radiation usage, and quantification of the reachable workspace (i.e., where these devices can access within the aneurysm).

During the first experiment, we compared the success rates, total procedure times, and radiation usage during the gate cannulation step. We did not find any statistical differences in any of the three metrics when comparing all the attempts. However, upon delving further into the effect of the position of the contralateral gate, we found that the steerable catheter achieved significantly higher success rates when the endograft was deployed in an unfavorable position. We also found that users required significantly less time cannulating the contralateral gate using the steerable catheter compared to the non-steerable catheter.

Our findings imply that, as expected, the improved steering offered by a steerable catheter can improve the outcome of these procedures in more challenging deployment positions and perhaps anatomies. During a procedure, a technical failure during this step often means an immediate conversion to open surgical repair since most devices on the market have no means of retracting a deployed endograft [30]–[34] these procedures have a greater perioperative mortality rate compared to EVAR and come with their own set of complications [35].

It was expected that the steerable catheter would also reduce the amount of time spent manipulating the device, which could mean fewer complications and a reduction in radiation exposure for both the interventionalists and the patients [36]. However, we were unable to observe overall statistical improvements in this aspect in our experiments. Procedure time reduction (and proportionally dose reduction is highly important. During some more challenging procedures, a previous study revealed that the interventionalists experienced more than 6 Gy of total radiation, with 2 Gy being enough to cause skin injury [37]. So, further improvements in steering technology may be warranted to improve in this aspect.

The analysis of the reachable workspace of the steerable catheters and non-steerable catheters provided further insight into the differences between the two device technologies. Although there were no statistical differences when looking at the entire cross-section of the aneurysm, when we divided the cross-section into different regions, we saw some differences in the area coverage for both catheters. The steerable catheter outperformed the non-steerable catheters in accessing the central region of the aneurysm, while the non-steerable catheters outperformed the steerable catheter along the periphery of the aneurysm. These differences were enhanced when analyzing the input of each individual non-steerable catheter compared to the steerable catheter. In almost every regional grouping, the steerable catheter outperformed each non-steerable catheter. Given the shorter time for using the steerable catheter and the fewer devices, this suggests an opportunity for increased procedure efficiency, despite the increased cost, with using the steerable catheter for more complex gate cannulations. This finding also re-amplifies the need for improved solutions that can potentially provide improved reachability and equal access to the entire anatomical target area.

Our findings were consistent with a previous study performed by our group that evaluated the workspace of a steerable catheter and non-steerable catheter in a peripheral vascular model [19]. The non-steerable catheter was unable to effectively access positions at different radii centered around a pivot point. In essence, the devices only allowed access to peripheral locations, whether it be within an aneurysm or the superficial femoral artery. The extra degree of freedom with a steerable catheter allows the user to change the radius at which they rotate, allowing more area coverage. So, when using conventional non-steerable catheters, surgeons may need differently shaped devices depending on the patient's anatomy. This can translate into more time-consuming manipulating devices, which risks greater complications and radiation exposure. Despite the benefits of a steerable catheter, there are still limitations to where the device can access. Factors such as the mechanical interactions between the catheter and the anatomy can still limit the reachable workspace.

As others have shown, anatomy has a significant impact on the challenges and success rates in EVAR [26]. The fundamental interactions between the anatomy and the catheter can limit their access, and the anatomy ultimately hinders any inputs from the surgeon. In addition, these procedures are performed under fluoroscopy, a 2D imaging modality. So, the surgeon does not know the true position of the device along the normal axis. To verify the position, operators may rotate the x-ray fluoroscopy source, but without simultaneous stereo imaging, any subsequent movements may affect the catheter's depth in the current perspective, and the issue of lack of depth perception persists. This challenge was especially apparent during the gate cannulation of position 2, where the contralateral gate was parallel to the ipsilateral limb. It was difficult to discern if the device had truly entered the contralateral gate due to the lack of depth perception, and users were unsure of how their manipulation affected the tip of the device. Such limitations motivate the design of new catheters. The presented study and methods can provide a systematic approach and baseline for comparing the performance of such new solutions to existing devices.

There are a few limitations to this study. First, we only used a few variations of non-steerable catheters and one variation of a steerable catheter. While generally, the devices used represent the technology that exists in terms of both groups, depending on the anatomy, an interventionalist may select a more specific catheter. They may also spend less time manipulating each catheter in favour of finding a catheter that works more efficiently. However, this study provides us with a baseline and overall quantitative characterization of these two device categories, which may still aid in the interventionalist's decision-making or permit further detailed analysis of a specific device in comparison to other devices in the future.

Next, the workspace analysis may not have fully represented various clinical scenarios. Even when dividing the entire cross-section into multiple regions, there are certain regions that the contralateral gate would possibly not reside in. Despite this, access to the less common regions can represent the challenge of more complex anatomies, demonstrating the flexibility in terms of access that may be required for these procedures.

In addition, we also had a relatively small sample size. This was limited due to the availability of resources. Some of our statistically insignificant findings may be associated with this limitation. However, we were still able to show trends that described the behavior of steerable and non-steerable catheters. Also, the presented method and protocol provide a systematic approach for obtaining such data for comprehensive studies when needed.

Finally, the phantom did not perfectly represent the corresponding anatomy and its mechanical properties. We used a rigid 3D-printed polymer created from a CT scan. For example, in vivo, when a catheter is within the iliac artery, it can straighten out and reduce the tortuosity, ultimately affecting the workspace and navigational capabilities of the device. There is also compliance in the artery, which affects the inputs from the operator. Furthermore, this model does not replicate the lubricity of the artery, which could reduce friction [23], [38]. However, these simplifications can potentially emphasize and standardize the interactions between the anatomy and the catheters.

This research can aid in developing and evaluating catheters that address the challenges of catheter control. For example, a device like the CathPilot or its side-looking version, called the SideEye, uses a system to create local steering unhindered by the shape of the anatomy, and evaluating such devices in similar experiments can demonstrate their benefits over conventional devices [39], [40].

In conclusion, this study comprehensively evaluates the performance characteristics of non-steerable and steerable catheters in the context of endovascular aneurysm repair (EVAR). Our systematic approach revealed distinct differences in the accessible workspaces each catheter type provides within the aneurysm. Specifically, while non-steerable catheters excel in peripheral regions, steerable catheters demonstrate superior performance in central regions, particularly under challenging anatomical conditions. These findings highlight the need for improved solutions that provide optimal performance regardless of target location. The study also suggests that the specific anatomical challenges of each EVAR case should guide the selection of catheter type. This insight could guide clinical decision-making, influencing catheter selection for optimal patient outcomes in EVAR procedures and directing future device design and development efforts for this application.

## CONFLICTS OF INTEREST

The authors have no conflicts of interest in relation to the material presented in this paper.